\documentclass[aps,prb,twocolumn,groupedaddress]{revtex4}
\usepackage{graphicx}
\usepackage{epsfig}
\usepackage{float}
                                                                                
\begin{document}

\title{Electronic Structure and Bulk Spin Valve Behavior in 
Ca$_3$Ru$_2$O$_7$}
\author{D.J. Singh$^1$ and S. Auluck$^2$}
\address{$^1$Condensed Matter Sciences Division, Oak Ridge
National Laboratory, Oak Ridge, TN 37831-6032, USA \\
$^2$Physics Department, Indian Institute of Technology, Roorkee
(Uttaranchal) 247667, India}

\begin{abstract}
We report density functional calculations of the magnetic properties
and Fermiology of Ca$_3$Ru$_2$O$_7$.  The
ground state consists of ferromagnetic bilayers, stacked antiferromagnetically.
The bilayers are almost but not exactly half-metallic. In the ferromagnetic
state opposite spin polarizations are found for in-plane and out-of-plane
transport. Relatively high out of plane conductivity is found for
the majority spin, which is relatively weakly conductive in-plane.
In the ground state in-plane quantities
are essentially the same, but the out of plane transport is strongly
reduced.
\end{abstract}

\maketitle
\date{\today }

The perovskite based ruthenates, $D_{1+n}$Ru$_n$O$_{3n+1}$, $D$=Sr,Ca
show a remarkable range of electronic and magnetic properties, even though
they are all based on Ru$^{4+}$
in octahedral environments with corner
sharing topologies. This includes robust itinerant
ferromagnetism (SrRuO$_3$) \cite{randall,kanbayashi},
paramagnetic ``bad" metals (CaRuO$_3$) \cite{cao1},
unconventional superconductivity (Sr$_2$RuO$_4$) \cite{maeno94}, an
antiferromagnetic (AFM) Mott insulator (Ca$_2$RuO$_4$)
\cite{nakatsuji}, and metamagnetic quantum
critical behavior (Sr$_3$Ru$_2$O$_7$)
\cite{grigera}.
This reflects exceptionally
strong dependence of electronic and magnetic properties
on lattice degrees of freedom, also seen in band structure
studies \cite{singh1,mazin1,sr-327,fang}.

The bilayer $n=2$ compounds
are of particular interest due to their borderline properties.
Sr$_3$Ru$_2$O$_7$ bridges the metallic ferromagnetic (FM) $n=\infty$ compound
SrRuO$_3$ and the paramagnetic Fermi liquid Sr$_2$RuO$_4$ and has
a metamagnetic quantum critical point \cite{grigera}
associated with borderline metallic ferromagnetism.
Ca$_3$Ru$_2$O$_7$ is intermediate between the bad metal
CaRuO$_3$ and the Mott insulator Ca$_2$RuO$_4$, and so may be
a useful window into the physics of clean materials near a
metal insulator transition.
Signatures of the borderline physics include
strong sensitivity of measured properties to minor sample variations,
observation of metallic like properties such as quantum oscillations
and finite linear specific heat coefficient in
non-conducting material,
a large lattice anomaly at the magnetic ordering temperature,
and low field metamagnetic transitions with
strong transport signatures
\cite{cao2,snow,cao3,cao4,cao5,mccall,karpus,ohmichi,cao6,lin,yoshida}.

Here, we report detailed electronic structure calculations for
Ca$_3$Ru$_2$O$_7$. These give insight into a number of
key observations -- magnetic structure, lattice
coupling, and transport. They were done
in the local spin density approximation (LSDA) using
the general potential linearized augmented planewave (LAPW) method with
local orbitals \cite{singh-book,singh-lo}. We used
the experimental low temperature (8K)
non-centrosymmetric $Bb2_1m$ crystal structure
\cite{yoshida}.
This is a centered structure, with two formula units, and a c(2x2)
in-plane cell doubling due to octahedral tilts and rotation.
The strong octahedral tilting differentiates this compound from
Sr$_3$Ru$_2$O$_7$.
Significantly, if insulating, this
non-centrosymmetric spacegroup would have a ferroelectric
polarization along $b$. This would be of interest in the
context of magnetoelectrics if a related well insulating, but magnetic,
material can be made.
The calculations were
highly converged with respect to basis set and zone samplings
\cite{mycode,WIEN}.
Approximately 2650 LAPW basis functions were used,
and twice that number for the doubled cell.
The Fermi surfaces are based on an interpolation
from 494 first principles points in the irreducible wedge of the zone,
except for the doubled AFM cell, where 194 points were used.

Ca$_3$Ru$_2$O$_7$ shows an AFM ordering at
$T_N$=56K and a transition from a metal to a low temperature
poorly conductive or insulating phase at 48K.
\cite{cao2}
Metamagnetic transitions to an effectively ferromagnetic phase,
with large changes in conductance occur at relatively low field,
depending on the field direction while magnetization measurements
imply that the RuO$_2$ layers themselves are ferromagnetically ordered,
with an AFM stacking \cite{mccall}.
Three possible AFM
stackings have been suggested -- ferromagnetic bi-layers stacked
antiferromagnetically, or bilayers that are internally antiferromagnetically
aligned, stacked either in a FM or an AFM fashion,
i.e. (1) $-UU-DD-UU-$, (2) $-UD-UD-UD-$ or (3) $-UD-DU-UD-$ along $c$.
Recent neutron scattering experiments favor ordering (1) over ordering (2),
based on observation of half order diffraction peaks. \cite{yoshida}
Considering the layered crystal structure and highly 2D electronic
structure (see below), magnetic interactions within a bilayer must
be very much stronger than inter-bilayer couplings.
Calculated energetics are given in Table \ref{e-tab}.
The non-spin-polarized
state (P) is unstable against ferromagnetism by 70 meV/f.u.,
but
state (2),
is favored over P by only 19 meV/f.u.
In plane antiferromagnetism
is also disfavored relative to ferromagnetism.
Thus AFM in the bilayers
is strongly disfavored, and the ground state must be (1) and not (3).
In fact, (1),
consisting of ferromagnetic bi-layers
stacked antiferromagnetically is
an additional 5 meV/f.u. lower than the FM state,
so the inter-bilayer coupling is weakly AFM.
We did
structural relaxations of all atomic positions in the $Bb2_1m$
cell. This yields an energy gain of 
only 16 meV (per 12 atom formula unit) for FM ordering,
supporting the experimental
crystal structure, and a larger 27 meV for the non-spin-polarized case,
the difference implying significant magnetoelastic coupling. For
magnetism with itinerant character, this can provide
an explanation of the structural anomalies at the ordering
temperature.

\begin{table}[tbp]
\caption{Calculated magnetic energies.
AF1 and AF2 are FM layers stacked
-UU-DD-UU- and -UD-UD-UD-, respectively. AFP is in-plane c(2x2) AFM.
}
\begin{tabular}{lccccc}
\\
\hline
ordering  & ~~P~~ & ~~FM~~ & ~~AFP~~ & ~~AF1~~ & ~~AF2~~ \\
\hline
E(meV/f.u.) & 0 & -72 & -52 & -77 & -19 \\
\hline
\end{tabular}
\label{e-tab}
\end{table}

Since the inter-bilayer coupling is weak, it is useful to start with the
FM state.
The calculations show that the
mechanism of moment formation and the major sheets
of Fermi surface are essentially the same as in the AFM ground state, as
are the density of states and in-plane Fermi velocities.
However, for the AFM case,
the majority and minority spin lie on top of each other
due to the alternating spin up and spin down bi-layers.

\begin{figure}
\vskip 0.2 cm
\epsfig{width=0.6\columnwidth,angle=270,file=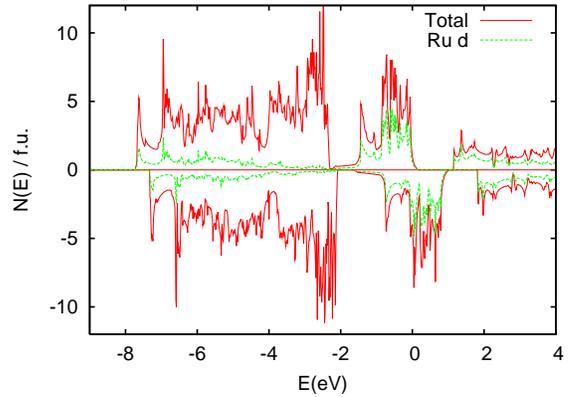}
\vskip 0.5 cm
\caption{(color online) Electronic density of states and projection onto
the Ru LAPW spheres, radius 2.05 $a_0$)
of FM Ca$_3$Ru$_2$O$_7$
on a per formula unit basis.
Majority spin is shown above the axis, and minority below.
}
\label{dos} \end{figure}

\begin{figure}
\vskip 0.2 cm
\epsfig{width=0.6\columnwidth,angle=270,file=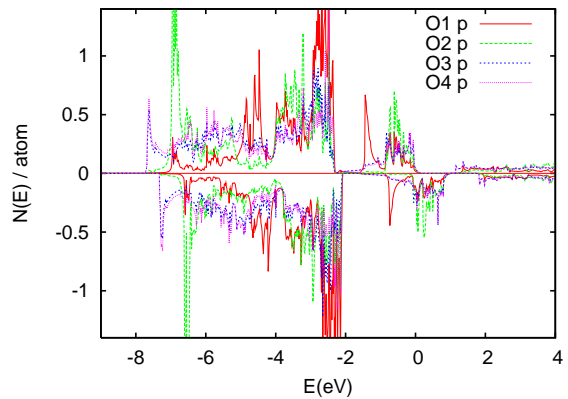}
\vskip 0.5 cm
\caption{(color online) Density of states projection onto the O LAPW
spheres (radius 1.65 $a_0$) on a per atom basis.
O1 is the apical O in the CaO rocksalt layers, O2 is the
bridging apical O joining the RuO$_2$ bilayers and O3 and O4
are the plane O in the RuO$_2$ layers.
}
\label{dos-o} \end{figure}

\begin{figure}
 \epsfig{width=0.65\columnwidth,angle=270,file=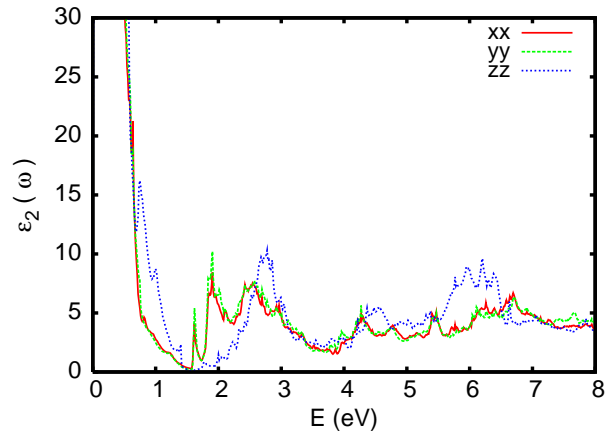}
\vskip 0.2 cm
\caption{Optical spectrum ($\epsilon_2$) for FM Ca$_3$Ru$_2$O$_7$.
}
\label{optics} \end{figure}

\begin{figure}
\epsfig{width=0.70\columnwidth,angle=270,file=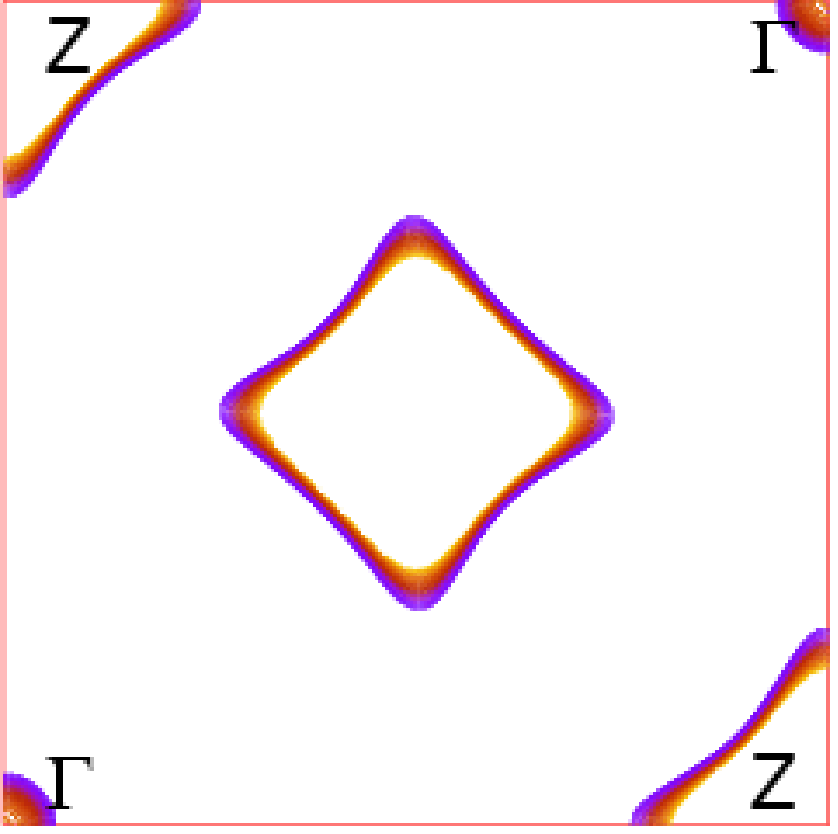}
\epsfig{width=0.70\columnwidth,angle=270,file=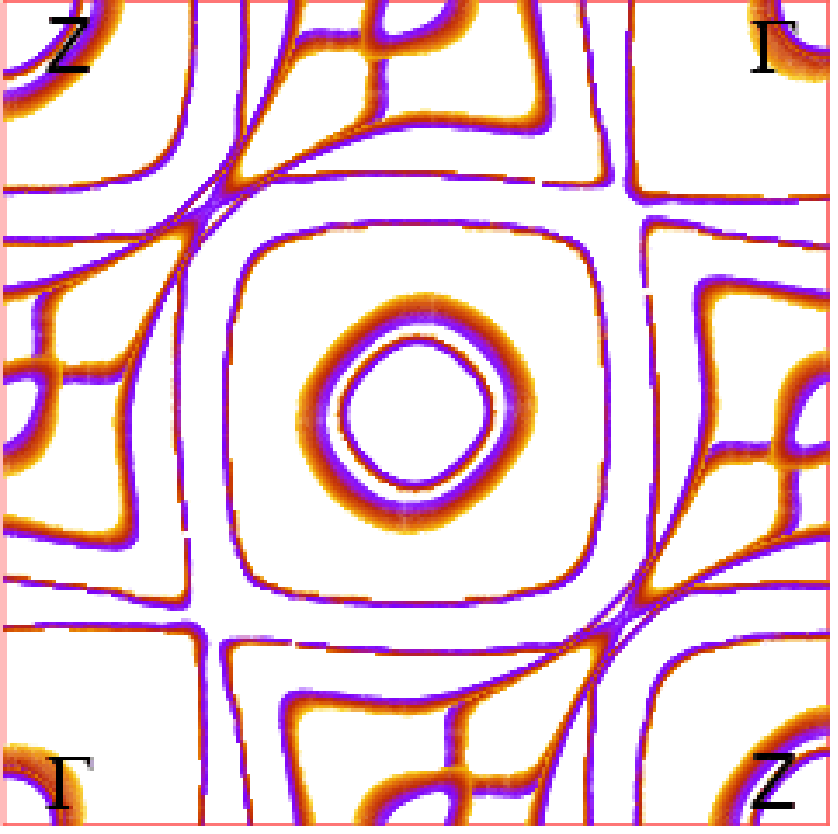}
\vskip 0.5 cm
\caption{(color online)
Majority (top) and minority (bottom) basal plane extended zone
FM Fermi surfaces.
The actual zone is half the area and
rotated 45$^\circ$. The plot shows a color map ranging from 1 mRy
below $E_F$ (blue) to 1 mRy above (red). The width is inversely
proportional to the velocity and electron (hole) surfaces
are blue (red) on the interior. The center is a $Z$ point;
alternating corners are $\Gamma$ and $Z$ points due to the
centered lattice.
}
\label{fs} \end{figure}

The spin
magnetization with FM ordering is 1.934 $\mu_B$/Ru, of
which only 1.23 $\mu_B$ is within the Ru LAPW sphere (radius 2.05 $a_0$).
The remaining $\sim 1/3$ is O in character,
comparable to other FM ruthenates \cite{mazin1,sr-327}.
The origin is the strong Ru-O hybridization in these $4d$ compounds,
combined with Hund's coupling on O.
The electronic density of states (DOS) and projections onto the Ru
and O LAPW spheres are shown in Figs. \ref{dos} and \ref{dos-o}.
The optical spectrum is in Fig. \ref{optics}
and the Fermi surface in Fig. \ref{fs}.
The low energy part of the optical spectrum below $\sim$ 1.5 eV, derives
from $d$-$d$ transitions, while the in-plane
peaks at $\sim$ 1.9 eV and $\sim$ 2.5 eV
are respectively
of majority and minority spin charge transfer $p$-$d$ character.
There is little in-plane anisotropy except at low energy (not shown).
Comparison with experiment
would be useful in testing the LSDA electronic structure.

Moment formation via an itinerant Stoner instability
depending on Ru-O hybridization,
similar to SrRuO$_3$ is clearly seen \cite{singh1,mazin1}. Thus
the O polarization plays a key role.
The importance of the Hund's coupling on O is also evident from the
exchange splitting of the O bands as seen in the DOS.
The spin moments within
the O LAPW spheres (radius 1.65 $a_0$) are 0.08 $\mu_B$ (apical O),
0.13 $\mu_B$ and 0.14 $\mu_B$ (two inequivalent in-plane O) and
0.21 $\mu_B$ (interlayer bridging O). These are reflected in the various
O contributions to the DOS near $E_F$ (Fig. \ref{dos-o}).
The large bridging O contribution, which also reflects the bonding of
the bilayer units, explains the strong interlayer
FM coupling.

Our results and the experimental
observation of relatively low field metamagnetism show that the
basic magnetic structure is ferromagnetic bilayers, with a very
weak inter-bilayer AFM coupling.
To understand the electronic structure we
begin with the FM bilayers.

Within an ionic model, each Ru has 4 $t_{2g}$ electrons, which
partially fill the $t_{2g}$ manifold. Because of band narrowing
due to the octahedral tilts and rotation in Ca$_3$Ru$_2$O$_7$,
this manifold is separated from the higher lying $e_g$ manifold
by a gap. If fully polarized, this would yield a spin moment
of 2 $\mu_B$/Ru, close to but larger than what we find.
The FM DOS shows that the
bilayers are almost, but not exactly half metallic. This is
similar to the colossal magnetoresistive (CMR) manganites
such as (La,Ca)MnO$_3$ \cite{pickett,nadgorny}.
In those materials, there is a finite density of states in both
spin channels, but $E_F$ falls near the minority spin
band edge, and those states are Anderson localized.
In FM Ca$_3$Ru$_2$O$_7$, $E_F$ falls very near
the majority spin band edge. However, because it is a clean
material, with no alloy scattering, there is no mechanism for
Anderson localization at low temperature. Thus, in contrast to
the CMR materials, as the temperature is lowered in FM
Ca$_3$Ru$_2$O$_7$ the majority spin channel will cross-over
from localized to metallic,
yielding an anisotropic partially spin polarized metal.

In a layered structure, the $t_{2g}$ manifold derives from two sets
of bands, $d_{xy}$ which is 2D, and $d_{xz}$ and $d_{yz}$, which
are 1D. For a bilayer, these split into
symmetric and antisymmetric combinations due to interlayer
hopping. Since the $d_{xy}$ is directed in-plane, its interlayer
coupling is expected to be smaller than for the $d_{xz}$ and $d_{yz}$
bands. Conversely, its in-plane band width may be expected to be
larger, following the number of hopping paths. This
qualitatively is what we find.

In Boltzmann theory
with the constant scattering time approximation, the conductivity varies
as $\sigma_{x} \sim N(E_F) v_{Fx}^2 \tau$, where $\tau$
is a scattering time. Both the paramagnetic
and FM electronic structures are quite anisotropic near
$E_F$ as expected from the layered crystal structure. This is also
apparent from the Fermi surfaces.
The exception is the majority spin in the FM case, which
has a low density of states and is 3D in character. This Fermi surface
has mixed character due to tilting, but is largely from
the bilayer antisymmetric combination of $d_{xy}$ orbitals,
Hopping through the CaO rocksalt layers is geometrically allowed by
octahedral tilts in this structure, so that even though the Fermi surface
has mostly $d_{xy}$ character it involves enough apical O character to produce
$c$-axis dispersion.

Turning to the minority spin channel, in addition to the highly 2D character,
there is also significant in plane anisotropy. In the constant
scattering time approximation, $b$ direction conductivity
is $\sim$50\% higher than along $a$.
This means that sample dependent orthorhombic twinning could
significantly affect the measured transport and should be considered
in interpreting experiments.
In plane conduction is dominated by
the minority channel and is highly spin polarized
\cite{mazin-sp},
$P_a^{(2)}$ = -0.88 and $P_b^{(2)}$= -0.91. On the other
hand, for the FM ordering, while $c$-axis conductivity is a factor of 20 lower,
it has a small opposite
polarization, $P_c^{(2)}$= +0.19.
The actual value of $P_c^{(2)}$ may actually be somewhat larger, because
the multi-sheet minority spin Fermi surface would provide more non-spin-flip
scattering
channels than the simple majority surface, and so the minority spin
scattering time may be shorter than the majority spin.
In any case, to our knowledge, this
type of direction dependent
sign of the polarization is unique to FM Ca$_3$Ru$_2$O$_7$. While here
it is a scientific curiosity, if found in materials that have
true ferromagnetic ground states and large $|P|$ for both directions,
it could be of practical importance for sources of spin polarized
electrons with spin depending on the contact geometry.

\begin{table}[tbp]
\caption{Fermi surface parameters of Ca$_3$Ru$_2$O$_7$ for the
P, FM and AFM ground state AF1. $N(E_F)$ is
given in eV$^{-1}$ per spin per f.u. Fermi velocities,
$<v_i^2>^{1/2}$ are in cm/s.}
\begin{tabular}{lcccc}
\\
\hline
  & $~~~N(E_F)~~~$ & $~~~~~v_a~~~~~$  & $~~~~~v_b~~~~~$  & $~~~~~v_c~~~~~$ \\
\hline
P            & 5.23     & 0.78x10$^7$ & 0.71x10$^7$ & 0.23x10$^7$ \\
F(majority)  & 0.95     & 0.64x10$^7$ & 0.71x10$^7$ & 0.58x10$^7$ \\
F(minority)  & 4.41     & 1.17x10$^7$ & 1.48x10$^7$ & 0.22x10$^7$ \\
AF1          & 2.76     & 1.0x10$^7$  & 1.4x10$^7$  & 0.10x10$^7$ \\
\hline
\end{tabular}
\label{fermi-tab}
\end{table}

The majority spin Fermi surface consists of a single highly corrugated
hole cylinder of mostly $d_{xy}$ character around the zone center,
which almost
pinches off at $k_z=0$.
The minority spin has several sheets of Fermi surface. The highest
velocity parts, which contribute most to the in-plane conductivity,
are from the symmetric and antisymmetric 1D
$d_{xz}$ and $d_{yz}$ sections. These reconnect to form
a square cylindrical $\Gamma$ centered section and an open section,
that forms from a square cylinder but opened at the zone boundary
at (1/2,0), which is (1/4,1/4) in the extended zone of Fig. \ref{fs}.
Along the other $b$-direction zone boundary there is a complex set
of Fermi surfaces. These derive from the 2D $d_{xy}$ sheets which
kiss the zone boundary at (0,1/2). These do not appear along the $a$
direction because the $d_{xy}$ sheet is gapped away by mixing with
the 1D $d_{xz}$ and $d_{yz}$ bands due to the octahedral tilts, which are
in the $a$-direction (i.e. around the $b$-axis). The repulsion from
the $d_{xz}/d_{yz}$ Fermi surfaces is strong enough to fold this sheet back
in the region around (1/2,1/2) leading to the complex structure seen around
this point.
The presence of the
$d_{xy}$ derived sections towards the $b$-direction zone boundary, but
not the $a$-direction, is the reason for the large
in-plane anisotropy of $\sigma$.
Besides the major sheets there are two smaller concentric nearly
circular cylindrical electron sections around $\Gamma$, which are also
of substantial $d_{xy}$ character. Significantly, these two minority spin
sheets have the same center and are intermediate in size between the maximum
and minimum size of the majority spin Fermi surface as a function of $k_z$.
However, they are mis-matched in that the majority surface contains holes
and the minority contains electrons.

Turning to the AFM ground state, as mentioned, one might expect
the electronic structure of the bi-layers to be essentially unchanged,
due to the weak inter-bilayer coupling,
and this is confirmed by the calculations. The only notable change
is a disruption of the $c$-axis transport.
There are no significant changes \cite{af-note} in $N(E_F)$ or the in-plane
$N(E_F)v_F^2$. However, the spin averaged $c$-axis $N(E_F)v_F^2$ is
strongly reduced by a factor of 10 relative to the FM case.
This is strong spin-valve physics.
Moreover, there is another factor.
For non-spin-flip scattering, which ordinarily dominates,
there are more scattering channels for the
minority spin due to its larger multisheet Fermi surface.
Thus with AFM stacking, there may be an important change in $\tau$,
which would further lower the $c$-axis conductivity, depending on
the relative importance of spin-flip and non-spin-flip scattering, which
in turn would depend on details of the sample and temperature.

In summery, we report calculations of the electronic and magnetic properties
of Ca$_3$Ru$_2$O$_7$, which elucidate the ground state ordering and
provide a framework for the strong connections between magnetic order
and transport properties, particularly at the metamagnetic transition.

We are grateful for helpful discussions with L. Balicas,
M. Braden, G. Cao, S.I. Ikeda,
D.G. Mandrus, I.I. Mazin, S. Nagler. We thank Y. Yoshida for a pre-publication
copy of Ref. \onlinecite{yoshida}
This work was supported by the U.S. Department of Energy.


\begin{references}
\bibitem{mycode}
Two codes were used. These were the author's LAPW code and
WIEN2K. These were cross-checked.
\end{references}
\end{document}